\documentclass[prl,letterpaper,twocolumn,tightenlines,superscriptaddress,       showpacs,preprintnumbers,nofootinbib]{revtex4}

\usepackage{amsmath,amssymb,amsfonts,graphicx,pifont}

\newcommand{\beq}{\begin{equation}}
\newcommand{\eeq}{\end{equation}}
\newcommand{\be}{\begin{eqnarray}}
\newcommand{\ee}{\end{eqnarray}}

\newcommand{\ms}{\Delta m^2}
\newcommand{\ts}{\sin^2 2\theta}

\newcommand{\dm}{\Delta m^2}

\begin{document}
\pagestyle{plain}

\preprint{VPI-IPNAS-10-10}

\title{Constraining sterile neutrinos with a low energy beta-beam}

\author{Sanjib Kumar Agarwalla}
\email{sanjib@vt.edu}

\affiliation{Department of Physics, IPNAS, Virginia Tech, 
Blacksburg, VA 24060, USA}

\begin{abstract}
We study the possibility to use a low energy beta-beam facility 
to search for sterile neutrinos by measuring the disappearance of electron
anti-neutrinos. This channel is particularly sensitive since it allows
to use inverse beta decay as detection reaction; thus it is free from
hadronic uncertainties, provided the neutrino energy is below the pion
production threshold. This corresponds to a choice of the Lorentz
$\gamma\simeq30$ for the $^{6}$He parent ion. Moreover, a
disappearance measurement allows the constraint of sterile neutrino
properties independently of any CP violating effects. A moderate
detector size of a few $100\,\mathrm{tons}$ and ion production rates of
$\sim2\cdot10^{13}\,\mathrm{s}^{-1}$ are sufficient to constrain mixing
angles as small as $\sin^22\theta=10^{-2}$ at 99\% confidence level.
\end{abstract}

\maketitle

\section{Introduction}
 
A large number of experiments have now convincingly demonstrated that
active neutrinos, {\it i.e.} left handed neutrinos which interact via
W and Z exchange, can change their flavor. More recently,
KamLAND~\cite{Abe:2008ee} is providing direct evidence for neutrino
oscillations. The oscillation of three\footnote{We know from the
invisible decay width of the Z boson~\cite{Yao:2006px}, that there
are 3 active neutrinos with $m<m_Z/2$.} active neutrinos describes
the global neutrino data very well, see {\it e.g.}~\cite{Maltoni:2004ei}. 
The fact that neutrinos oscillate implies that at least two of the mass 
eigenstates have a non-zero mass.  Most models to accommodate massive neutrinos 
introduce right handed neutrinos, {\it i.e.} states which do not couple to W 
and Z bosons. These right handed neutrinos can either directly provide a
Dirac mass term or they mediate the seesaw
mechanism~\cite{Minkowski:1977sc}. In the latter case, the right
handed neutrino tends to be very heavy with
$m_R\sim10^{12}-10^{15}\,\mathrm{GeV}$.  However, in the most general
scenario there is a 6x6 mass matrix whose entries are essentially
unknown. To obtain the physical neutrino masses at scales below the
electroweak phase transition this mass matrix has to be diagonalized
and its eigenvalues are the neutrino masses relevant for low energy
observations. Since the entries of the mass matrix are not known it
can gives rise to any spectrum of eigenvalues and thus neutrino
masses. The need to describe the oscillation of three active neutrinos
implies that \emph{at least} three eigenvalues have to be of
$\mathcal{O}(1\,\mathrm{eV})$ or less. However, there can be $0-3$
additional small eigenvalues corresponding to light neutrino states,
which due to the Z decay width bound would have to be sterile.  Thus,
we conclude that sterile neutrinos are theoretically well motivated by
the observation of neutrino mass.  Furthermore, we see that there can
be one or more sterile neutrinos which are light enough to play a role
in neutrino oscillations. Note, that all these considerations are
entirely independent of any experimental claims to have seen sterile
neutrinos, like, for example, the one by
LSND~\cite{Athanassopoulos:1997pv}.

In this work we explore how well sterile neutrinos can be constrained
by a dedicated oscillation experiment based on a low energy beta-beam facility.
For other physics which can be studied using a low energy beta-beam,
{\it e.g.} see~\cite{volpeadd}. The proposed experiment is a
disappearance experiment and will be sensitive to
$\dm=0.5-50\,\mathrm{eV}^2$ and can probe mixing angles as small as
$\sin^22\theta=10^{-3}-10^{-2}$. The oscillation probability for
our purposes is a two flavor $\bar\nu_e\rightarrow\bar\nu_s$
oscillation and the survival probability of $\bar\nu_e$ is given by
{\small{
\begin{equation}
\label{eq:probability}
P_{\bar e \bar e}=1-\sin^22\theta \sin^2\frac{\Delta m^2 L}{4 E}\,.
\end{equation}
}}

The most sensitive experiments in this mass range looking for the
disappearance of $\bar\nu_e$ have been the reactor neutrino
experiments: Bugey~\cite{Declais:1994su} and
Chooz~\cite{Apollonio:2002gd}. The Bugey bound is valid in the range
$\dm=0.1-1\,\mathrm{eV}^2$ and it can constrain $\sin^22\theta$ below
$5\cdot10^{-2}$.  The bound from Chooz is
$\sin^22\theta\simeq10^{-1}$ for $\dm>0.01\,\mathrm{eV}^2$.

\section{Experimental setup}

The concept we propose here is based on using a pure $\bar\nu_e$ beam
from the beta decay of completely ionized radioactive ions circulating
inside a storage ring. The Lorentz boost $\gamma$ of these ions will
be chosen such that the resulting $\bar\nu_e$ have an energy below the
pion production threshold. In this case, the by far most likely
reaction is inverse beta decay on free protons.  In this experiment we
have an accurate theoretical understanding of the neutrino flux,
spectrum and cross section. The detector will be placed so close to
the neutrino source that oscillation will happen \emph{within} the
detector itself. Thus the different parts of the detector will
effectively act as near and far detector. This allows the cancellation
of most systematical errors in a similar fashion as in modern reactor
neutrino experiments.

\subsection{Beta-beam}

In~\cite{zucc}, the idea of using beta decay of exotic nuclei to
produce well defined neutrino beams was introduced. The basic
observation is, that if a beta decaying nucleus is moving with a
Lorentz $\gamma\gg 1$, the isotropic neutrino emission will be collimated 
into a beam. The isotope we consider here is $^6$He, which beta decays 
with a half life of $0.81\,\mathrm{s}$ and has an end-point energy
$E_0=4.02\,\mathrm{MeV}$. We assume that $3\cdot 10^{12}$ ions are
injected per second into the storage ring and the beta-beam complex 
operates for $1.6\cdot10^{7}\,\mathrm{s}$ per calendar year and our experiment 
runs for a total of five years.

We consider a magnetic field $B=5\,\mathrm{T}$ inside the storage ring. 
Taking $\gamma=30$ this yields magnetic radius $\rho=56\,\mathrm{m}$ for 
the $^6_2$He$^{++}$ ion. The useful fraction, $f$, of decays then is given by
the ratio of the length of the straight section, $S$, to the overall
circumference of the ring, $f=\frac{S}{2S+2\pi\rho}$.
With $S=10\,\mathrm{m}$, we obtain $f=2.7\%$.

\begin{figure}[t]
\includegraphics[width=\columnwidth]{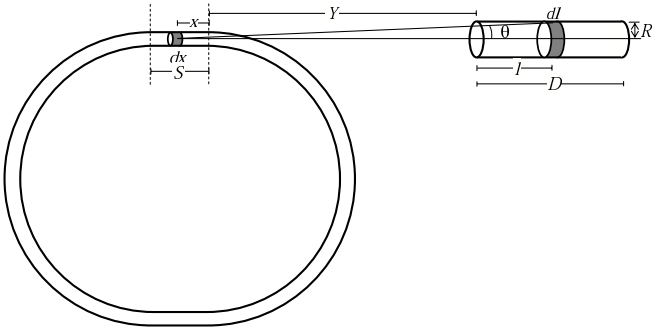}
\caption{\label{fig:set-up} Schematic of the detector and storage ring.}
\end{figure}

\subsection{Event rate calculation}

Figure~\ref{fig:set-up} shows a schematic of the setup we consider.
In this scheme we will have a cylindrical detector whose symmetry axis
is aligned with the straight section of the storage ring. The free
parameters are (see figure~\ref{fig:set-up}): the length of the
straight section ($S$), the distance between the front end of the
storage ring and the front end of the cylindrical detector ($Y$), the
radius ($R$) and the length ($D$) of the detector. In the following, we
will derive a general expression for the neutrino event rate as a a
function of these free parameters.

\begin{figure}[h]
\includegraphics[width=\columnwidth]{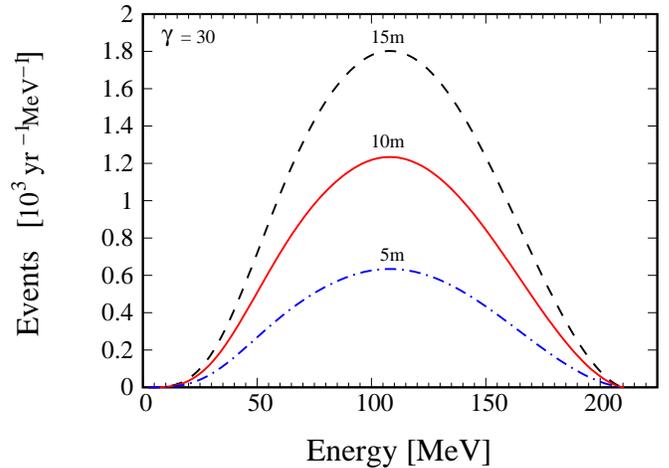}
\caption{\label{fig:unosc} The un-oscillated event rate as a
function of neutrino energy.  The event rate has been calculated
including all geometrical effects and with a luminosty of
$3\cdot10^{12}\,\mathrm{ions}\,\mathrm{s}^{-1}$. The different lines
show the result for different values of the length of the straight
section, $S$, as indicated by the labels next to each line.}
\end{figure}

Neglecting the small Coulomb corrections to the
beta-spectrum\footnote{We checked that these corrections are
negligible for our purposes.}, the lab frame neutrino beta-beam flux
per unit length of the straight section in units of
$\mathrm{sr}^{-1}\,\mathrm{MeV}^{-1}\,\mathrm{s}^{-1}\,\mathrm{m}^{-1}$
emitted at an angle $\theta$ with the beam axis is described by
{\footnotesize{
\beq
\phi^{Near}(E,\theta)
 =\frac{1}{4\pi}\frac {g} {m_e^5 \,f}
\frac{1}{\gamma(1-\beta \cos\theta)}
 (E_0 - E^*) E^{*2} \sqrt{ (E_0-E^*)^2-m_e^2},
\label{eq:near_flux}
\eeq
}}
where $m_e$ is the electron mass, $E_0$ is the electron total
end-point energy and $E^*$ is the rest frame energy of the emitted
neutrino\footnote{Quantities without the `$\ast$' refer to the lab
  frame.}.  $f$ is the phase space factor associated with the beta
decay of the nucleus. $\gamma$ is the Lorentz boost such that
{\small $E^* = \gamma E(1-\beta \cos\theta)$}, $E$ being the neutrino 
energy in the lab frame. $g \equiv N_0/S$ is the number of useful decays 
per unit time per unit length of the straight section.

To calculate the resulting number of events in a cylindrical detector
of radius $R$ and length $D$ aligned with the beam axis it is
necessary to integrate over the length $S$ of the straight section of
the storage ring and the volume of the detector. Here we assume that
the beam is perfectly collinear and has no transverse
extension\footnote{Note, that the beam size is of order
  $10^{-2}\,\mathrm{m}$, whereas all other length scales are of order
  $\sim10\,\mathrm{m}$.}. The un-oscillated event rate in a detector
placed at a distance $Y$ from the storage ring is given by
{\footnotesize{
\beq
\frac{dN}{dt}
=n \varepsilon \int_0^S \: dx \int_0^D \: d\ell
\int_0^{\theta'} d\theta \:{2\pi} \sin\theta
\int_{E_{min}}^{E^{\prime}} \: dE \:\phi^{Near}(E,\theta)\:
\sigma(E),
\eeq
}}
where {\small $\tan\theta^{'}(x,\ell)=\frac{R}{Y+x+\ell}\,~~{\rm
and}~~E^{\prime}=\frac{E_0-m_e}{\gamma(1-\beta\cos\theta)}$}.

Note, that the baseline, relevant for oscillations, is $L = Y+x+\ell$.
Here, $n$ represents the number of target nucleons per unit detector
volume, $\varepsilon$ is the detector efficiency which is taken to be
unity in our calculation. $E_{min}$ denotes the energy threshold for
our detection method.  We work with a threshold of $25\,\mathrm{MeV}$
which ensures that our events are well above the backgrounds.
$\sigma(E)$ stands for the inverse beta decay reaction cross section
\cite{cross}, which is the predominant reaction channel at the
considered energies. Figure~\ref{fig:unosc} shows the resulting
un-oscillated event rates for $S=10\,\mathrm{m}$, $Y=50\,\mathrm{m}$,
$R=3.6\,\mathrm{m}$ and $D=28.7\,\mathrm{m}$. This corresponds to a
detector mass of $1\,\mathrm{kton}$.

\subsection{Optimization of the geometry}

\begin{figure}
\includegraphics[height=.24\textheight]{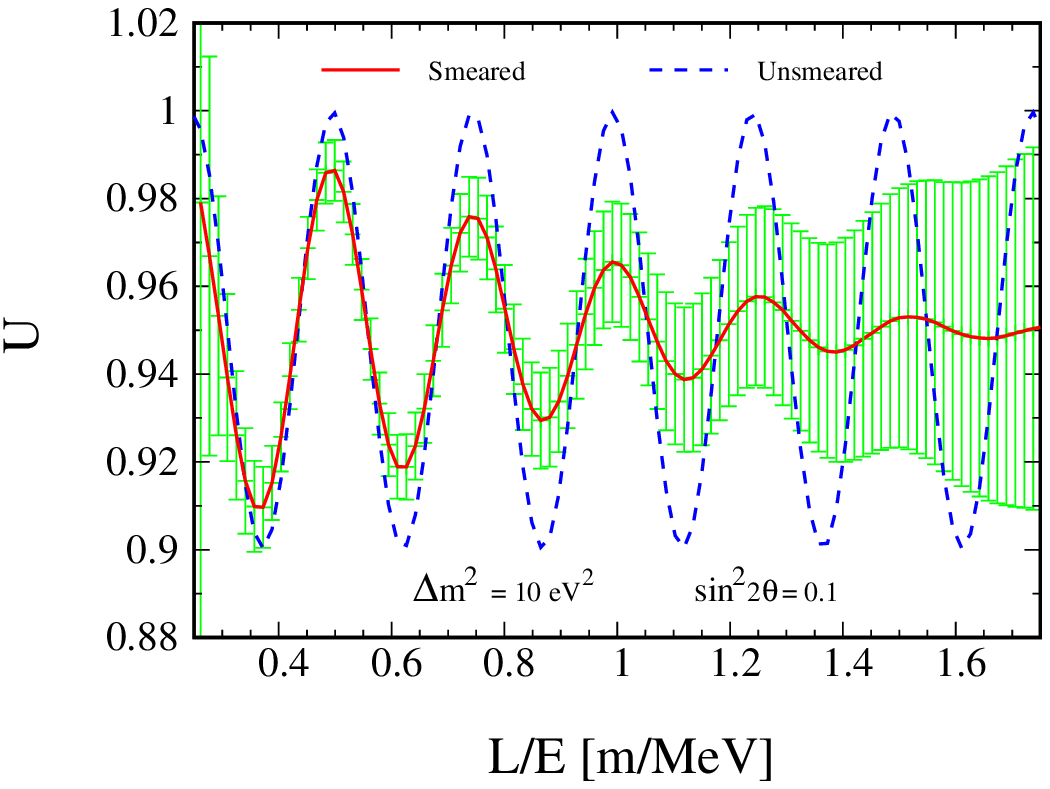}
\includegraphics[height=.24\textheight]{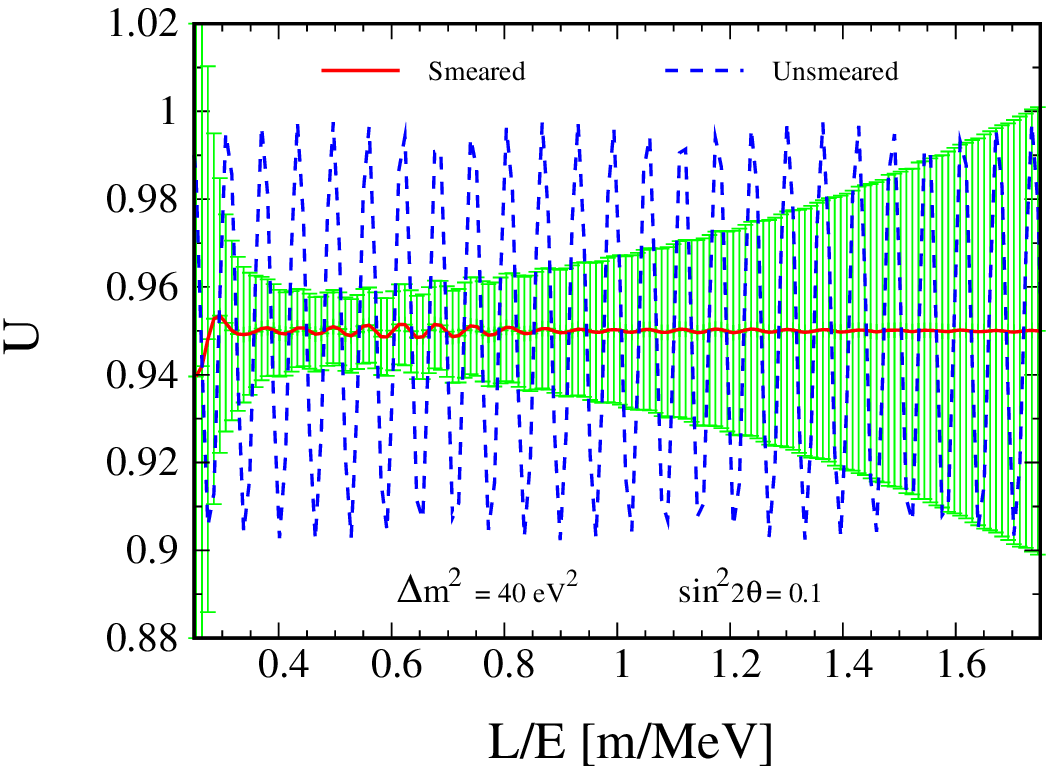}
\caption{\label{fig:oscillation} This figure shows the ratio of
oscillated to un-oscillated events as a function of the
reconstructed $L/E$.  In upper panel, $\ms$ = 10 eV$^2$ and in lower
panel $\ms$ = 40 eV$^2$. The value of mixing term $\ts = 0.1$. The
red (solid) line includes all geometrical effects and the detector
resolution, whereas the blue (dashed) line assumes a point source of
neutrinos.}
\end{figure}

The goal is to obtain an experimental configuration which has optimal
sensitivity to the disappearance of $\bar\nu_e$ corresponding to a
mass squared difference $\Delta m^2=1-10\,\mathrm{eV}^2$. For this
optimization we fixed the detector mass at $1\,\mathrm{kton}$, thus $D$
is entirely determined by $R$ or {\it vice versa}. We tested the 
values $Y=\{30,50,70,90\}\,\mathrm{m}$. We studied variations in
$\gamma$ from $20-35$.  This range was chosen to stay below or close
to the pion production threshold. We also changed the detector radius from
$3.5-4.5\,\mathrm{m}$.  We found that within those options the
configuration with $\gamma=30$, $S=10\,\mathrm{m}$,
$L=50\,\mathrm{m}$ and $D=28.7\,\mathrm{m}$ is optimal.

Using these numbers, figure~\ref{fig:oscillation} shows the resulting
ratio of oscillated to un-oscillated event rates for two different
vales of $\Delta m^2$ as a function of the reconstructed $L/E$.  The
blue line assumes that the neutrinos are all generated at in one point
in the middle of the straight section. The red line fully accounts for
all the geometry effects. Clearly, for $\Delta m^2=10\,\mathrm{eV}^2$
(upper panel), several oscillation periods can be resolved. A
comparison between the amplitudes of those periods allows to cancel
systematics to a large extent (see also, figure~\ref{fig:sys}). At
$\Delta m^2=40\,\mathrm{eV}^2$ (lower panel) only an average
suppression can be observed and the sensitivity is entirely determined
by the achievable systematic errors.

\subsection{Detector}
\label{sec:detector}

The detector we envisage is essentially similar to the MiniBooNE
detector~\cite{AguilarArevalo:2008qa}, however with a cylindrical
shape. The important background will all be beam based, since beam-off
backgrounds will be well measured and cosmic ray events can be
tagged with high efficiency as was done in similar near surface
experiments~\cite{AguilarArevalo:2008qa}.  The beam energy has been
selected to be below threshold for pion production, therefore there
are few channels available for neutrino interactions.  Only charged
current quasi-elastic scattering on carbon and electron elastic
scattering can mimic the signal of the inverse beta decay primary
positron.  These event types will experience the same disappearance
rates due to oscillations, but the neutrino energy reconstructed under
the inverse beta decay hypothesis would be systematically less than
the true neutrino energy.  At these energies the cross sections for
these background interactions are smaller by at least an order of
magnitude, so any effect they might have on the measured oscillation
parameters, in particular $\sin^22\theta$ would be at most 10\%, and
they can be accounted for and corrected.

To further reduce the impact of these beam based backgrounds, the
detector should be optimized for the detection of inverse beta decay.
Typically this is done by tagging the primary positron with the free
neutron capture in delayed coincidence with a mean lifetime of $\sim
200\,\mu\mathrm{s}$ in undoped organic scintillator. However,
observing the $2.2\,\mathrm{MeV}$ gamma ray from neutron capture on
hydrogen can be a significant challenge in a detector tuned to see
events in the 50 to $200\,\mathrm{MeV}$ range.  Additionally, the long
delay time will put several beam bunches between the primary and
secondary events and will therefore increase the probability of false
tags.  Adding gadolinium to the scintillator would help somewhat by
increasing the tag energy to $8\,\mathrm{MeV}$ and reducing the
capture time to $\sim 30\,\mu\mathrm{s}$.  Nevertheless, even if the
neutron tag was highly efficient, at least some of the quasi-elastic
events on carbon will also have a correlated neutron capture tag.
Another possibility is to design the detector to be sensitive to the
positron direction.  We expect the elastic scattering events to be
peaked in the very forward direction, while the quasi-elastic carbon
events will have a much broader angular distribution.  The angular
distribution of the hydrogen inverse beta decay events will fall
somewhere in between.  Sensitivity to the angular distribution can be
achieved by reducing the scintillation light to a point where
\v{C}erenkov light can be distinguished.

To achieve an energy resolution of 10\% or better over the 50 to
$200\,\mathrm{MeV}$ range should be possible with a photo-cathode
coverage of 10\% if we assume approximately equal parts \v{C}erenkov
and scintillation light~\cite{AguilarArevalo:2008qa}.  Position and
timing resolutions, needed to correlate events with beam bunches,
should be achievable at the half meter and $10\,\mathrm{ns}$ level.
With a bunch spacing of $100\,\mathrm{ns}$ or greater, this resolution
would provide sufficient space between bunches to demonstrate the
rejection of non-beam backgrounds. At this level of detector
resulution the $L/E$ uncertainty is fully dominated by the unkown
production point in the straight section of the decay ring.

\section{Results}
\label{sec:results}

\begin{figure}[!t]
\includegraphics[height=.24\textheight]{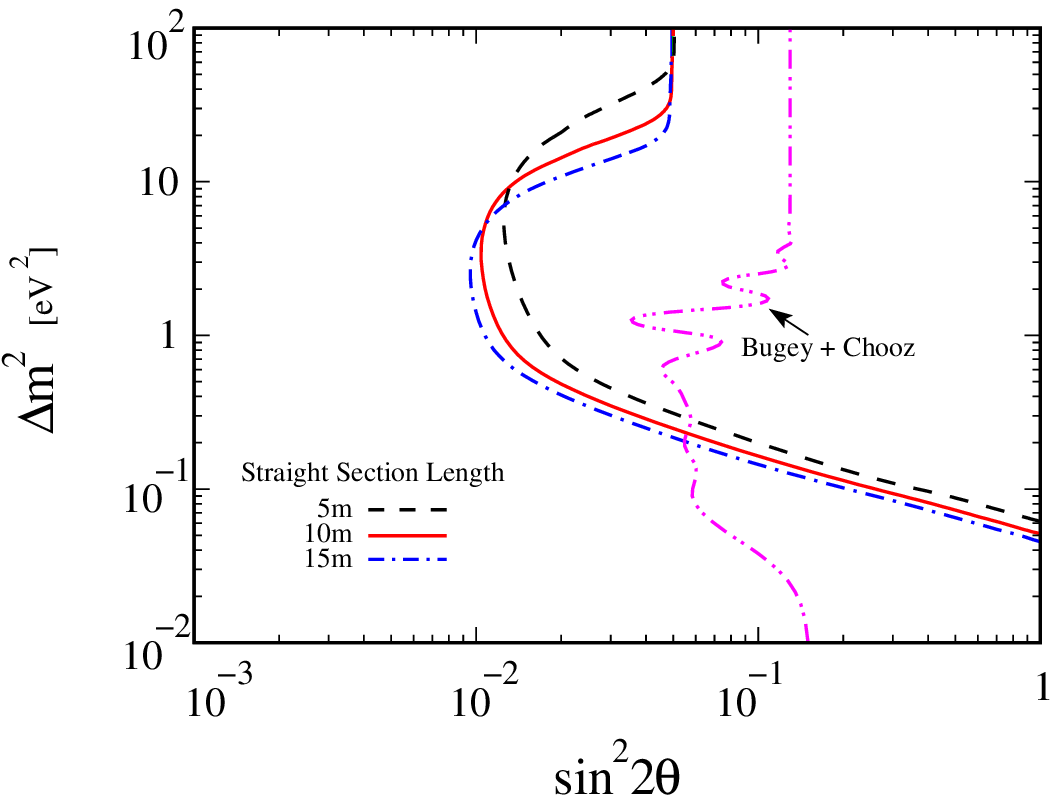}
\includegraphics[height=.24\textheight]{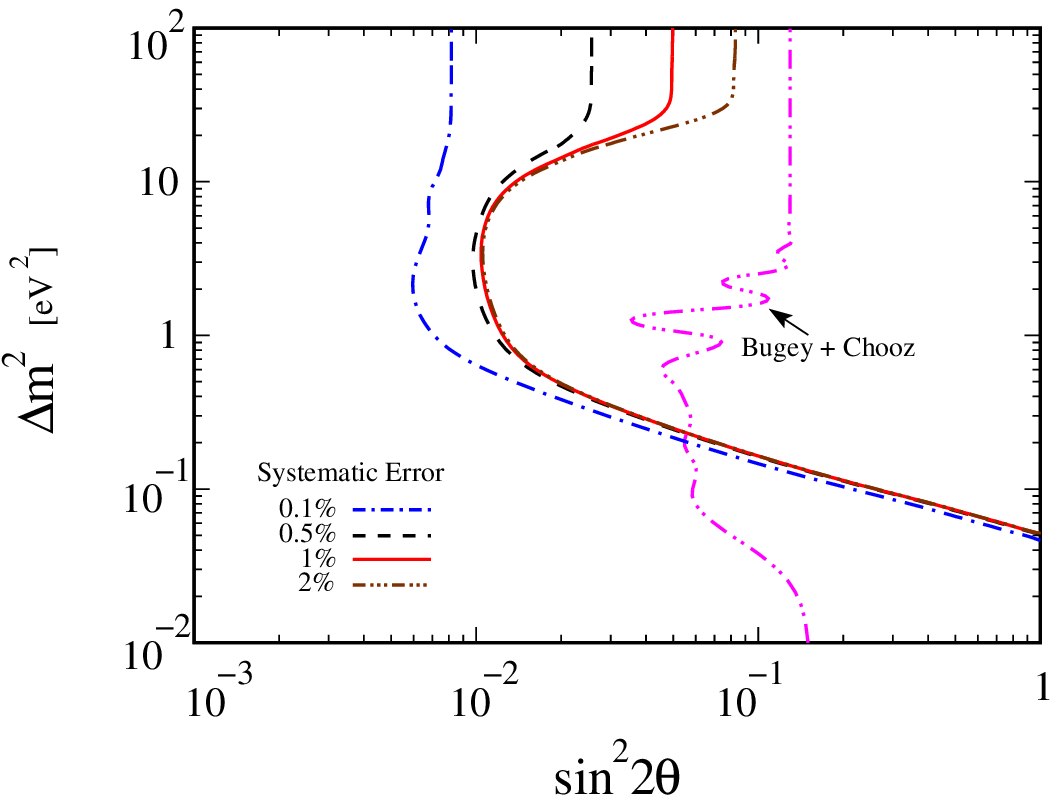}
\caption{\label{fig:sys} Exclusion plots of sensitivity to
active-sterile oscillations in a near detector low energy beta-beam
set-up. These are 99\% CL limits (1 dof). In upper panel, we vary the
straight section of the storage ring while in the lower panel, we
show the performance taking various values of expected systematic
error of the considered setup. The pink (dash-dot-dotted line) line
shows the current, combined limit on $1-P_{\bar e\bar e}$ from
Bugey~\cite{Declais:1994su} and Chooz~\cite{Apollonio:2002gd}. }
\end{figure}

Figure~\ref{fig:sys} shows the obtainable sensitivity in the
$\sin^22\theta$-$\Delta m^2$ plane at 99\% CL. Our default
configuration is shown as red solid line in both panels.  The setup
proposed here improves on the existing limit for $\Delta
m^2\geq0.2\,\mathrm{eV}^2$. In the range $1\,\mathrm{eV}^2<\Delta
m^2<10\,\mathrm{eV}^2$ the improvement is one order of magnitude or
better. The upper panel shows how the sensitivity changes with
varying the length of the straight section. A longer straight section
(dash-dotted line) implies a large fraction of useful decays and thus
better statistics.  At the same time the $L/E$ resolution is reduced.
As a result the sensitivity extends to smaller mixing angles (higher
statistics) but at smaller $\Delta m^2$ (resolution). The analogous
arguments holds also for a shorter straight section (dashed line),
which yields smaller statistics but better resolution. This improves
sensitivity to $\Delta m^2>10\,\mathrm{eV}^2$. Thus the length of the
straight section can effectively be used to tune the experiment to the
desired range of $\Delta m^2$ and it can be envisaged to have running
periods with different straight section lengths within the same setup.
The lower panel shows variations of the systematic error. 
Again our default setup with systematic error of $0.01$ is shown 
as red, solid line. At very small value of systematic error of $0.001$,
the sensitivity (dash-dotted blue line) becomes essentially
independent of $\Delta m^2$, once the first oscillation maximum can be
observed. Note, that this very small value of systematic error of $0.001$ 
probably is not attainable in a real experiment. For more realistic values 
of systematic error in the range $0.005-0.05$, the sensitivity limit does not
change for $\Delta m^2\leq 10\,\mathrm{eV}^2$, which is due to the
cancellation of the normalization error between different oscillation
maxima. This is also illustrated in figure~\ref{fig:oscillation}.

\begin{figure}[!t]
\includegraphics[height=.24\textheight]{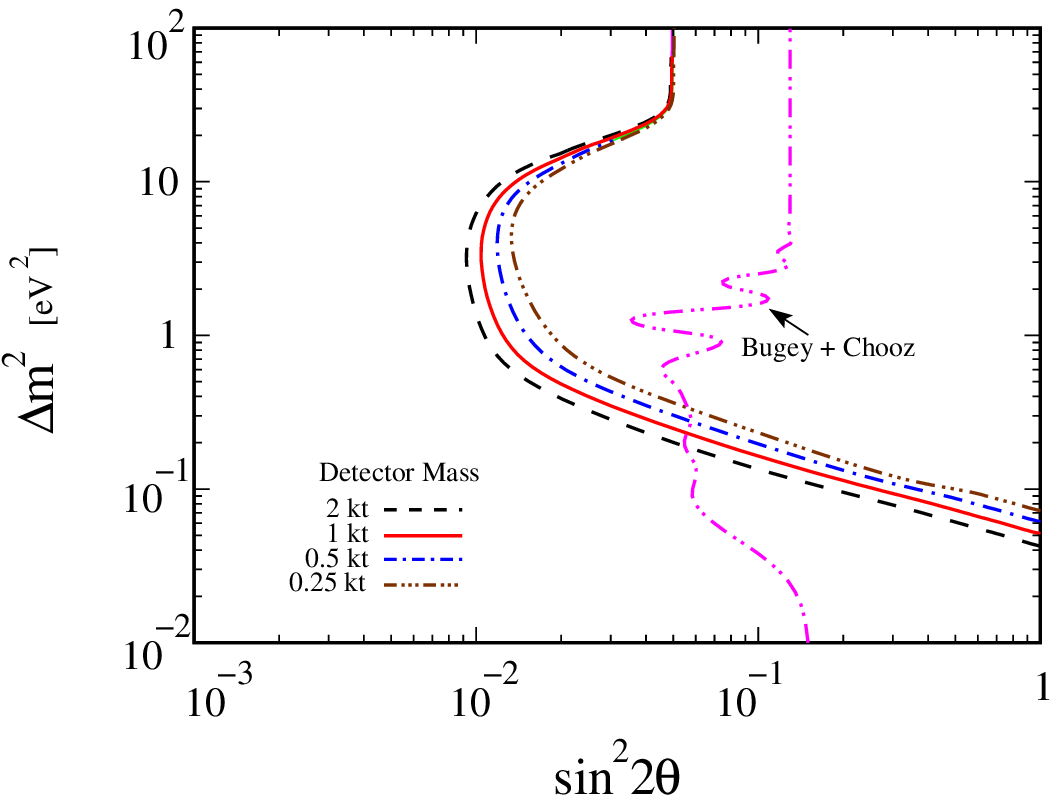}
\includegraphics[height=.24\textheight]{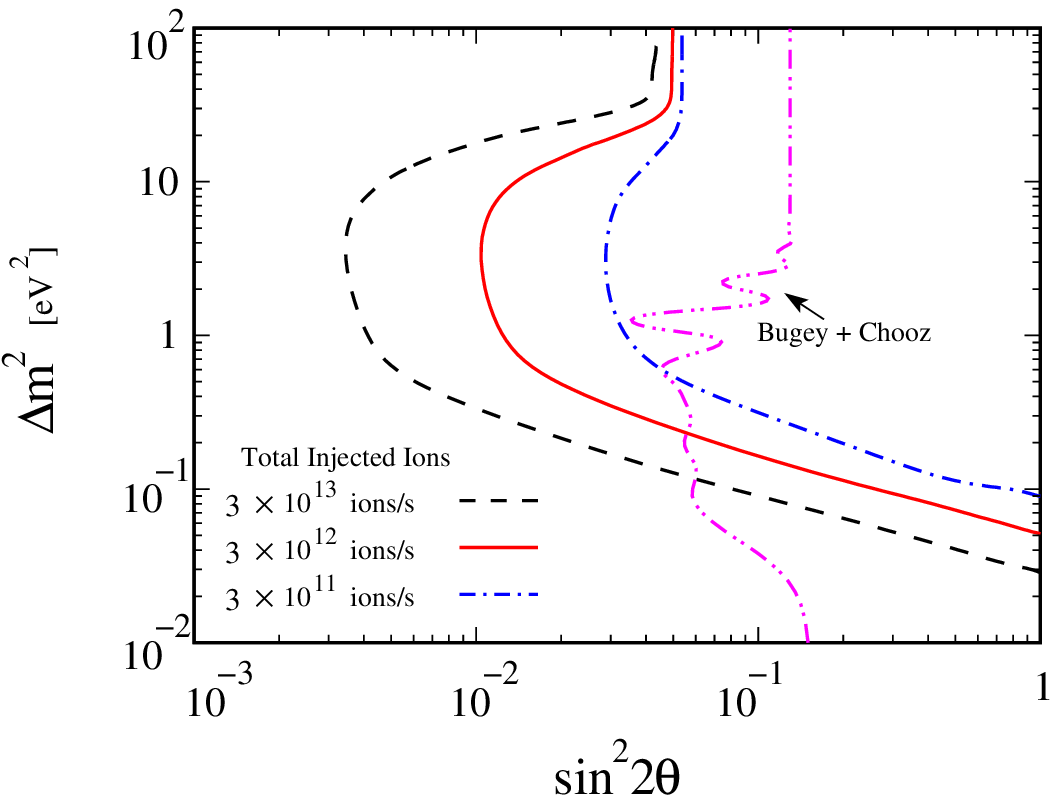}
\caption{\label{fig:luminosity} Exclusion plots of
sensitivity to active-sterile oscillations in near detector low
energy beta-beam set-up. These are 99\% CL limits (1 dof). In upper
panel, we choose different values of detector mass in the range 0.25
to 2 kton. Lower panel depicts the performance of the set-up for
different values of total injected ions. The pink
(dash-dot-dotted) line shows the current, combined limit on 
$1-P_{\bar e\bar e}$ from Bugey~\cite{Declais:1994su} and
Chooz~\cite{Apollonio:2002gd}.}
\end{figure}

Figure~\ref{fig:luminosity} shows the obtainable sensitivity in the
$\sin^22\theta$-$\Delta m^2$ plane at 99\% CL. Our default
configuration is shown as red solid line in both panels.  The upper
panel shows how a variation of detector mass by a factor of 10
changes the sensitivity.  Remarkably, this change is quite small: the
detector density and aspect ratio are fixed. Thus with increasing
detector mass, the additional detector mass will be exposed to a
weaker neutrino flux for purely geometrical reasons. The lower
panel shows the change of sensitivity for a changing beam luminosity
and here the effect is quite pronounced as every additional ion
contributes equally to improve the sensitivity. Note, the our default
beam luminosity of $3\cdot10^{12}\,\mathrm{ions}\,\mathrm{s}^{-1}$ is
based on existing technology and thus may be somewhat conservative.

\section{Conclusions}
\label{sec:conclusions}

We have studied a near detector setup at a low-$\gamma$ beta-beam
facility for its ability to constrain the disappearance of electron
anti-neutrinos for mass squared differences $\Delta
m^2=1-10\,\mathrm{eV}^2$. The key point is, that for a suitably chosen
geometry several oscillation maxima occur within the same detector and
thus a disappearance measurement at the sub-percent level becomes
possible without requiring a stringent control on systematic errors.
We focused on using a beam from the decay of $^6$He, which produces
electron anti-neutrinos. This allows to use inverse beta decay as
detection reaction and we can exploit the well defined relationship
between the positron and neutrino energy. Thus, we have a very clean
sample of electron anti-neutrino events.  We carefully optimized the
geometry and beam energy and found that $\gamma=30$ yields the best
sensitivity while still having the bulk of neutrinos below the pion
production threshold. In order to have sufficient resolution in $L/E$
we had to reduce the length of the straight section down to
$10\,\mathrm{m}$, which makes this setup unique. Note, that this
allows our experiment to run parasitically in a low energy beta-beam
facility since we use only around 3\% of all ions. For a conservative
beam luminosity of $3\cdot10^{12}\,\mathrm{ions}\,\mathrm{s}^{-1}$ and
detector mass of $1\,\mathrm{kton}$ we obtain a sensitivity to
$\sin^22\theta\simeq10^{-2}$ (99\% CL) for $\Delta
m^2=1-10\,\mathrm{eV}^2$.



\end{document}